\documentclass[]{spie}  %>>> use for US letter paper
\usepackage{amsmath,amsfonts,amssymb}
\usepackage{graphicx}
\usepackage[colorlinks=true, allcolors=blue]{hyperref}

\usepackage{graphicx}
\usepackage{multirow}
\usepackage{multicol}
\usepackage{adjustbox}
\usepackage{epsfig}
\usepackage{amsmath}
\usepackage{amssymb}
\usepackage{mwe}
\usepackage{acro}
\usepackage{amssymb}
\usepackage{xcolor,colortbl}
\usepackage{tabularx}
\usepackage{relsize}
\usepackage{pifont}
\usepackage{booktabs} 

\title{GLAM: Glomeruli Segmentation for Human Pathological Lesions using Adapted Mouse Model}

\author[a]{Lining Yu}
\author[b]{Mengmeng Yin}
\author[a]{Ruining Deng}
\author[a]{Quan Liu}
\author[a]{Tianyuan Yao}
\author[a]{Can Cui}
\author[d]{Yitian Long}
\author[b]{Yu Wang}
\author[e]{Yaohong Wang}
\author[c]{Shilin Zhao}
\author[b]{Haichun Yang}
\author[a,b,d]{Yuankai Huo*}
\affil[a]{Department of Computer Science, Vanderbilt University, Nashville, TN, USA}
\affil[b]{Department of Pathology, Microbiology and Immunology, Vanderbilt University Medical Center, Nashville, TN, USA}
\affil[c]{Department of Biostatistics, Vanderbilt University Medical Center, Nashville, TN, USA}
\affil[d]{Data Science Institute, Vanderbilt University, Nashville, TN, USA}

\affil[e]{Department of Anatomical Pathology, UT MD Anderson Cancer Center, TX, USA}

\authorinfo{Corresponding author: Yuankai Huo: E-mail: yuankai.huo@vanderbilt.edu}

\begin{document} 
\maketitle

\begin{abstract}
Moving from animal models to human applications in preclinical research encompasses a broad spectrum of disciplines in medical science. A fundamental element in the development of new drugs, treatments, diagnostic methods, and in deepening our understanding of disease processes is the accurate measurement of kidney tissues. Past studies have demonstrated the viability of translating glomeruli segmentation techniques from mouse models to human applications. Yet, these investigations tend to neglect the complexities involved in segmenting pathological glomeruli affected by different lesions. Such lesions present a wider range of morphological variations compared to healthy glomerular tissue, which are arguably more valuable than normal glomeruli in clinical practice. Furthermore, data on lesions from animal models can be more readily scaled up from disease models and whole kidney biopsies. This brings up a question: ``\textit{Can a pathological segmentation model trained on mouse models be effectively applied to human patients?}" To answer this question, we introduced GLAM, a deep learning study for fine-grained segmentation of human kidney lesions using a mouse model, addressing mouse-to-human transfer learning, by evaluating different learning strategies for segmenting human pathological lesions using zero-shot transfer learning and hybrid learning by leveraging mouse samples. From the results, the hybrid learning model achieved superior performance. 
\end{abstract}


\section{INTRODUCTION}
\label{sec:intro}  % \label{} allows reference to this section

Recent advancements such as ~\cite{isensee2021nnu, huang2017densely,dolz2018hyperdense} in image processing and deep learning have significantly influenced the enhancement of computer-assisted analysis of medical images , particularly in the quantification of animal tissues via WSI. Within the domain of renal pathology, which is characterized by its unique challenges in pathology image analysis, the glomerulus is identified as critical functional units essential for clinical assessments, according to ~\cite{jiang2021deep,han2023fastcellpose,ostergaard2020automated}. 

Considerable research including~\cite{janowczyk2016deep, komura2019machine} has investigated the use of deep learning in the quantification of glomeruli within the field of renal pathology, employing technologies such as convolutional neural networks (CNNs) like~\cite{gadermayr2019cnn, esteva2019guide, wang2019pathology, kamnitsas2017efficient}. Interestingly, for glomerulus segmentation task, mouse-to-human transfer learning has been proven to be practical by~\cite{souza2023mouse, ostergaard2020automated}.
Nonetheless, these studies like~\cite{yang2022glomerular} have devoted relatively little attention to the specific area of glomerular lesions. Glomerular lesions, which signify glomerular damage, serve as primary markers for various renal pathologies, according to~\cite{saikia2023mlp}. Nonetheless, the segmentation of glomerular lesions, owing to their minuscule size and the variability encountered in intra- and inter-observer analyses, presents a more formidable challenge than the segmentation of healthy glomeruli. The analysis of glomerular lesions in humans is a fundamental aspect of nephrology, offering critical insights for diagnosis, understanding disease mechanisms, guiding treatment decisions, assessing prognosis, preventing kidney failure, and facilitating research into new treatments. Moreover, very few, if any, publicly accessible datasets are available for training segmentation models, which include a broad spectrum of lesion types in mouse kidney pathology images. 

\begin{figure*}[bth]
\begin{center}
 \includegraphics[width=0.8\linewidth]{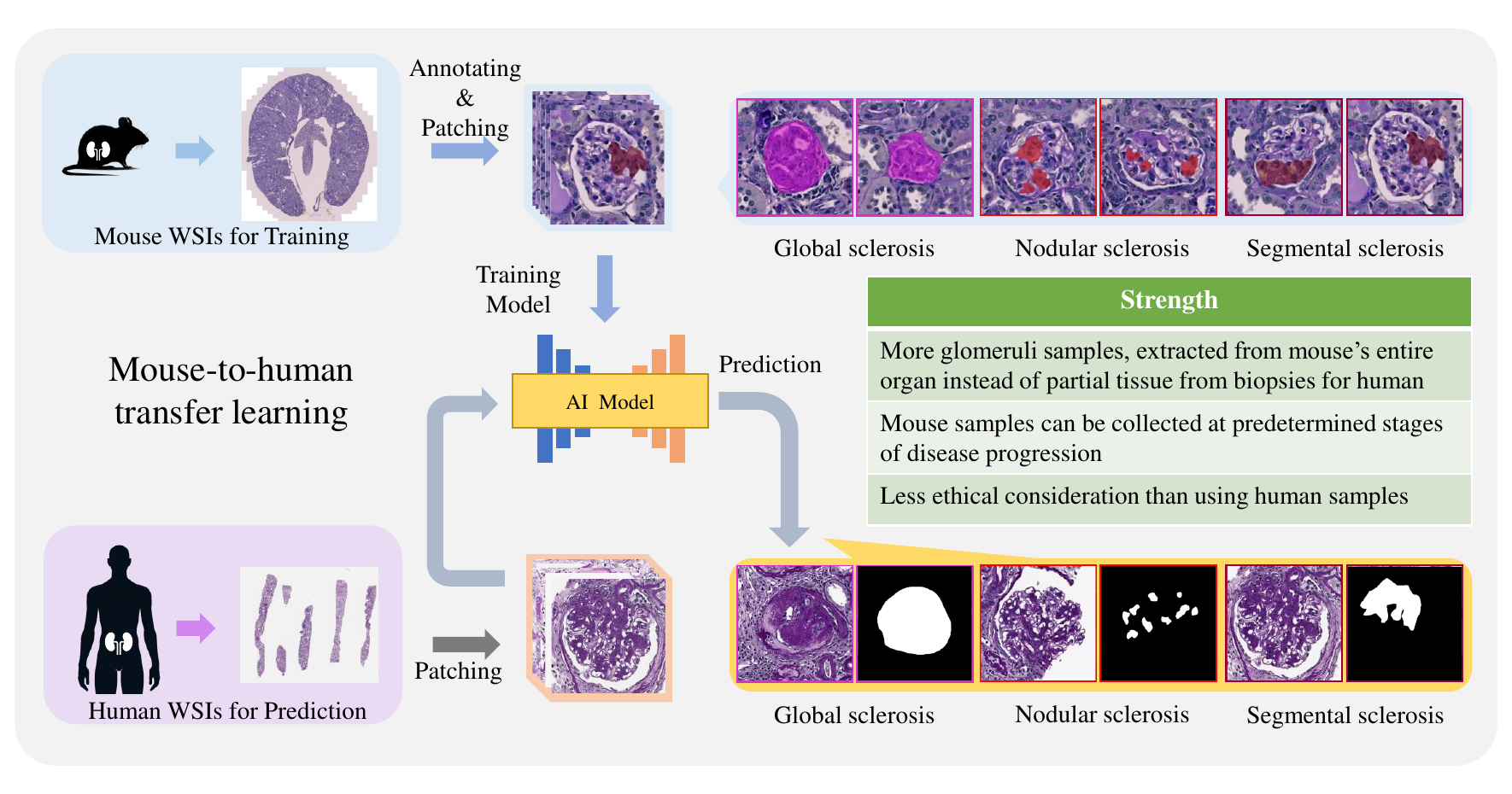}
\vspace{-0.5cm}
\end{center}
   \caption{\textbf{Overview of the learning framework.} This figure provides an overview of transfer learning for glomerular segmentation from mouse to human, highlighting the strenth of using mouse samples compared to human samples.}   
  \label{fig:Overview}
\end{figure*}

\begin{figure*}
\begin{center}
\includegraphics[width=1.0\linewidth]{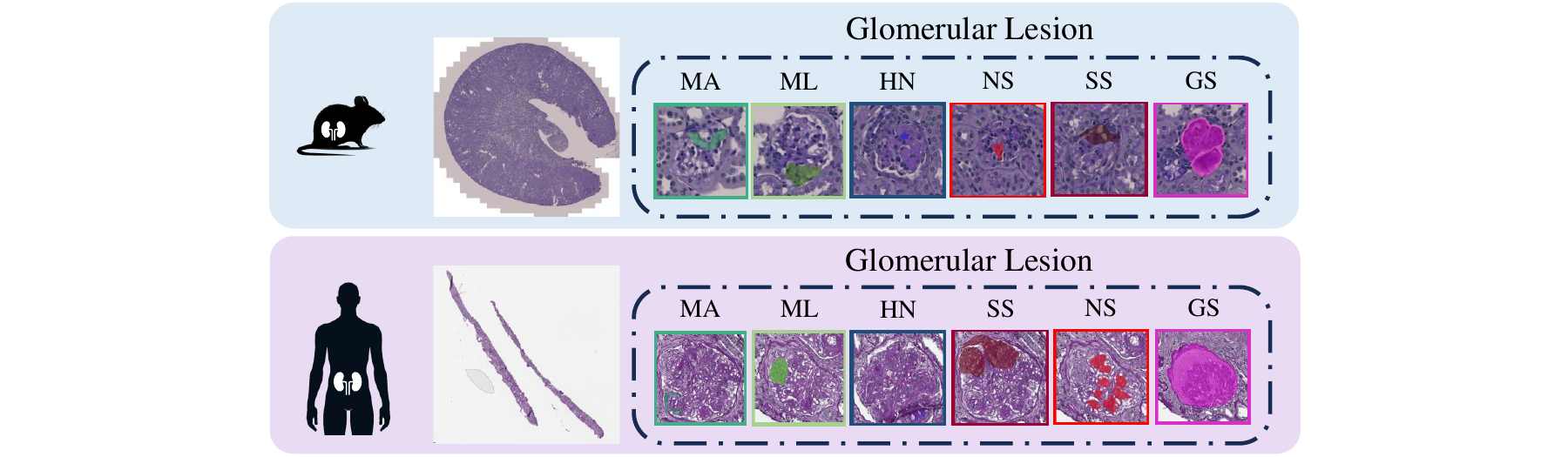}
\vspace{-0.5cm}
\end{center}
   \caption{\textbf{Lesion Classes.} This figure shows fine-grained classes of mouse and human glomerular images that are stained with PAS. Glomeruli with lesions: global sclerosis(GS), Hyalinosis(HN), mesangial lysis(ML), microaneurysm(MA), nodular sclerosis(NS), and segmental sclerosis(SS).}
   \label{fig:types}
\end{figure*}

To address the issue of costly human sample acquisition, we employ model trained by mouse samples to make prediction on human aiming to isolate the specific lesion type from a given section containing glomeruli, as shown in Fig.\ref{fig:Overview}. Additionally, our dataset stands out as it includes a wider variety of lesion types (illustrated in Fig.\ref{fig:types}) compared to those used in previous studies.

Our contributions in this paper are summarized as follows:

$\bullet$ We introduce GLAM, a deep learning framework designed for the fine-grained segmentation of human kidney lesions through the adaptation of a mouse model. 

$\bullet$ We present GLAM zero-shot transfer learning, which effectively utilizes knowledge from mouse samples to predict human outcomes.

$\bullet$ We implement the GLAM transfer learning strategy by adapting a mouse model for human kidney lesion segmentation. This approach leverages a combination of readily available mouse samples and human samples, enhancing the model's ability to generalize and accurately predict lesion types in human data.

\section{Method}

\begin{figure*}[t]
\begin{center}
\includegraphics[width=1.0\linewidth]{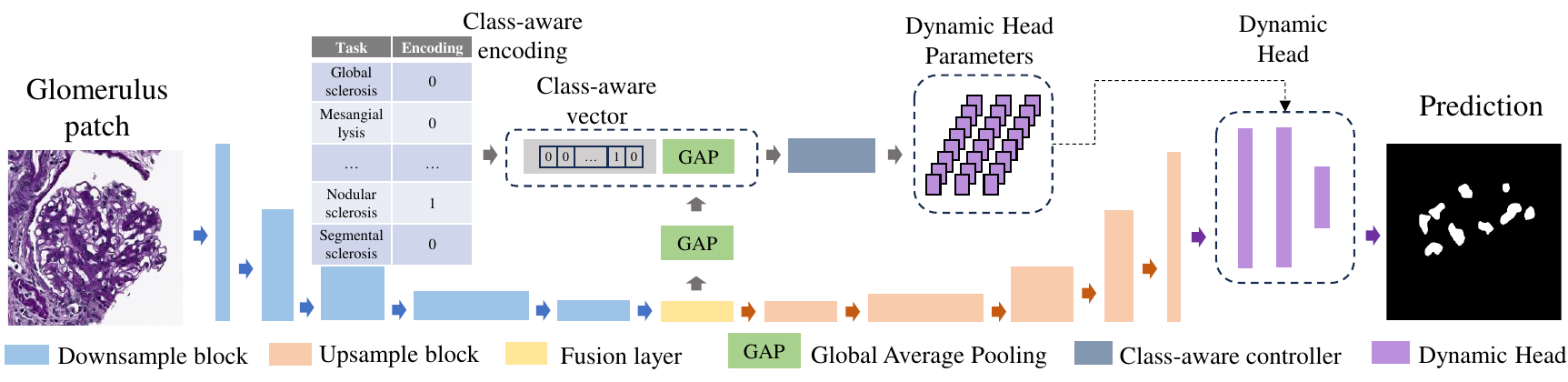}
\end{center}
\vspace{-0.5cm}
\caption{{\textbf{GLAM pipeline.} This figure introduces the network structure of the proposed dynamics head. Specifically, it contains a residual U-Net backbone, a class-aware controller, and a dynamic segmentation head.}}
\label{Framework}
\end{figure*}

We introduce the GLAM network, inspired from the research by~\cite{deng2022single}, tailored for our glomerular lesion segmentation tasks. GLAM Zero-shot network is a unified segmentation network that leverages a residual U-Net architecture to segment various lesion classes within partially labeled images for pathology analysis. The comprehensive structure of the GLAM Zero-shot network is detailed in Figure referred to as Fig.\ref{Framework}.

Unlike from fully labeled dataset, each image included in partially labeled dataset contains the annotations of only a specific class of glomerular lesion. We assign each segmentation on each class of lesion as a task for task-awareness for task-awareness, by encoding the task as a $m$-dimensional one-hot vector by~\cite{chen2017fast}. The $m$ is the number of lesion classes, The encoding calculation for $T_k$, a class-aware vector of $i$th class of lesion is shown as follows:
\begin{equation}
T_k = \left\{
\begin{array}{ll}
1, & \text{if } k = i \\
0, & \text{\textit{otherwise}}
\end{array}
\right. \quad k = 1,2,...,m
\end{equation}

Dynamic filter generation was introduced by \cite{zhang2021dodnet} to generate the kernels specialized to a particular class of lesion segment tasks. The image feature $F$ is aggregated by a global average pooling (GAP) with $T_k$. The kernel parameters $\omega$ are computed as follows:

\begin{equation}
\omega = \phi(\text{GAP}(F)||T_k; \Theta_\phi)
\end{equation}

\noindent where $\Theta_\phi$ represents the controller parameters, ``$||$" represents the concatenation operation to combine high-level image features and the class-aware vector.$\phi$ reprensents a task-specific controller with a single 2D convolutional layer. 

The dynamic head was designed to achieve multi-label segmentation with three layers, denoted by $\omega_1$, $\omega_2$, $\omega_3$. The predictions of lesions can be generated as follows:

\begin{equation}
P = ((((M * \omega_1) * \omega_2) * \omega_3))
\end{equation}

\noindent where $ * $ is convolution, $M$ is the output from the decoder.

\begin{table*}[b!]
% \begin{tabular}{llllllll}
%\centering
\caption{Data Collection}

\begin{center}
\vspace{0.5cm}
%\begin{adjustbox}{width=0.8\textwidth}
\begin{tabular}{l|cc|ccccccc}
\toprule
\multirow{2}{0.5in}{Samples} & \multirow{2}{0.3in}{Size} & \multirow{2}{0.3in}{Scale}  & \multicolumn{7}{c}{\#Patch}\\
\cmidrule(lr){4-10}
&   &   & GS & HN & ML & MA & NS & SS & Sum\\
\midrule
Mouse  & $512^{2}$ & $80\times$ &  346 & 187 & 73 & 198 & 328 & 193 & 1325 \\
Human  & $512^{2}$ & $40\times$ & 370 & 36 & 23 & 57 & 229 & 72 & 787 \\
\bottomrule
\end{tabular}
%\end{adjustbox}
\end{center}
\label{table:data}
\end{table*}

\section{Experiments}

\subsection{Data description}

The datasets utilized in this study, as detailed in Table.\ref{table:data}, encompass both a mouse glomerular lesion dataset and a human glomerular lesion dataset. The mouse dataset collect patches from whole-slide images (WSIs) obtained from 17 mouse kidneys, spanning 6 lesion classes. The human dataset consists of patches derived from WSIs of 19 patients, also spanning 6 lesion classes. Initially, each patch in both datasets measured 1024x1024 pixels, but they were subsequently resized to 512x512 pixels and each patch contains at least one type of lesion. Based on the inherent size differences between human and mouse glomeruli. Human glomeruli are generally larger than those of mice, so to achieve a similar level of detail and ensure consistent visualization of lesions across species, we conducted our analysis at different magnifications: 40x for human samples and 80x for mouse samples, ensuring similar visualization and assessment of the lesions.

\begin{table*}[h!]
% \begin{tabular}{llllllll}
%\centering
\caption{Performance of different models trained with mouse samples on human glomerular lesion segmentation. Dice similarity coefficient (\%, the higher, the better), Hausdorff Distance (Micron unit, the lower, the better), and Mean Surface Distance (micron unit, the lower, the better) are used for evaluation, where VM refers to model selection utilizing validation results from mouse data, whereas VH indicates model selection based on validation results from human data. The bold mark indicates the best performance.}

\begin{center}
\begin{adjustbox}{width=1.0\textwidth}
\begin{tabular}{l|ccccccccccccccc}
\toprule
\multirow{2}{0.8in}{Method} & \multicolumn{3}{c}{Global sclerosis} & \multicolumn{3}{c}{Hyalinosis} & \multicolumn{3}{c}{Mesangial lysis} & \multicolumn{3}{c}{Microaneurysm}\\
\cmidrule(lr){2-4}
\cmidrule(lr){5-7}
\cmidrule(lr){8-10}
\cmidrule(lr){11-13}
& Dice & HD & MSD & Dice & HD & MSD	 & Dice& HD & MSD
 & Dice & HD & MSD\\
\midrule
U-Nets Zero-shot & 70.6 & 185.0 & 51.6 & 49.9 & 352.4 & 317.0 & 49.2 & 521.6 & 441.8 & \textbf{50.9} & 273.5 & 119.7 \\
DeepLabV3s Zero-shot  & 78.2 & 164.5 & 47.3 & \textbf{52.7} & 351.7 & 282.1 & 52.4 & 359.1 & 280.6 & 49.9 & 340.3 & 268.2\\
Multi-class Zero-shot & 78.0 & 143.5 & 40.2 & 49.9  & 391.0 & 355.7 & 49.2 & 521.6 & 441.8 & 49.4 & 422.9 & 347.8 \\
Swinunetr Zero-shot& 81.5 & 147.6 & 37.3 & 50.1  & 319.0 & 251.5 & 49.3 & 225.1 & \textbf{94.2} & 49.5 & 255.6 & \textbf{109.1} \\
Segmenter Zero-shot& \textbf{82.5} & 146.6 & 36.6 & 50.4 & \textbf{267.8} & \textbf{224.7} & 49.2 & 521.6 & 441.8 & 50.7 & 251.6 & 121.2 \\
\bottomrule
GLAM Zero-shot(VM) & 79.3 & 137.2 & 39.8 & 49.9  & 389.4 & 348.6 & \textbf{57.8} & \textbf{244.8} & 118.5 & 50.5 & \textbf{240.9} & 122.2 \\
GLAM Zero-shot(VH) & 82.0 & \textbf{125.3} & \textbf{35.9} & 49.9 & 387.5 & 351.9 & 56.2 & 248.9 & 96.8 & 50.2 & 243.8 & 111.4 \\
\bottomrule
% GLAM MH2H & 94.1 & 76.3 & 14.3 & 57.8 & 181.5 & 85.5 & 54.6 & 284.7 & 213.5 & 51.0 & 241.7 & 162.4 \\
% \bottomrule
\end{tabular}
\end{adjustbox}
%\begin{adjustbox}{width=1.0\textwidth}
\begin{tabular}{l|ccccccccc}
\toprule
\multirow{2}{0.8in}{Method} & \multicolumn{3}{c}{Nodular sclerosis} & \multicolumn{3}{c}{Segmental sclerosis}  & \multicolumn{3}{c}{Average}\\
\cmidrule(lr){2-4}
\cmidrule(lr){5-7}
\cmidrule(lr){8-10}
& Dice & HD & MSD & Dice & HD & MSD	 & Dice& HD & MSD\\
\midrule
U-Nets Zero-shot  & 67.8 & 169.4 & 62.8 & 47.8 & 492.8 & 399.9 & 56.0 & 332.5 & 232.1 \\
DeepLabV3s Zero-shot  & 56.4 & 287.3 & 166.5 & 60.8 & \textbf{156.2} & \textbf{55.2} & 58.4 & 276.5 & 183.3 \\
Multi-class Zero-shot & 48.4  & 490.3 & 371.2 & 47.8 & 492.8 & 399.9 & 53.8 & 410.4 & 326.1 \\
Swinunetr Zero-shot & 53.3  & 231.3 & 105.8 & 49.6 & 179.1 & 63.4 & 55.6 &\textbf{ 226.3} & \textbf{110.2} \\
Segmenter Zero-shot & 49.4 & 251.1 & 141.1 & 49.4 & 261.1 & 117.7 & 55.3 & 283.3 & 180.5 \\
\bottomrule
GLAM Zero-shot(VM) & 67.8  & 172.6 & 61.3 & 69.5 & 190.6 & 59.0 & 62.5 & 229.3 & 124.9 \\
GLAM Zero-shot(VH) & \textbf{69.3}  & \textbf{160.7} & \textbf{52.3} & \textbf{69.8} & 194.0 & 64.1 & \textbf{62.9} & 226.7 & 118.7 \\
\bottomrule
% GLAM MH2H & 76.8 & 145.9 & 37.0 & 70.1 & 200.9 & 63.5 & 67.7 & 152.8 & 43.3 & 67.4 & 183.4 & 88.5 \\
% \bottomrule

\end{tabular}
%\end{adjustbox}
\end{center}
\label{table:M2H_result}
\end{table*}

\subsection{Implementation details}

We designed the image pool by incorporating strategies from Cycle-GAN by \cite{zhu2017unpaired} for each category, with a batch size of four and an image pool size of eight. Excess images in the pool, when exceeding the batch size, were removed and input into the network.

To evaluate the model's performance, we used metrics such as the Dice similarity coefficient (Dice), Hausdorff distance (HD), and Mean Surface Distance (MSD). The best-performing model on the validation dataset was selected based on the average Dice score over 200 epochs. This model's performance was then assessed using the same metrics. All experiments were conducted on a 24G NVIDIA RTX A5000 GPU.

To evaluate the zero-shot model's performance, we first used it in a mouse-to-human (M2H) scenario, predicting human samples with a model trained on mouse data. After training, its effectiveness was assessed using the human testing dataset. In a separate evaluation, we assessed the model's performance in predicting human outcomes without mouse sample data. For this, the model was exclusively trained on a human dataset and then evaluated on the human testing dataset in a human-to-human (H2H) scenario. Lastly, we predicted human samples using models trained on a hybrid dataset comprising both mouse and human samples. The performance of this model in a mouse and human-to-human (M\&H2H) scenario was evaluated using the human testing dataset.

We compared the introduced network to baseline models, including: (1) multiple individual U-Net models (U-Nets) by ~\cite{ronneberger2015u}, (2) multiple individual DeepLabv3 models (DeepLabv3s) by ~\cite{lutnick2019integrated}, and (3) a multi-class segmentation model for partially
labeled datasets by ~\cite{gonzalez2018multi} for renal pathology quantification. Additionally, the performance of the network was evaluated against transformer baselines (4) a hybrid neural network architecture that combines the Swin Transformer with the U-Net transformer encoder for enhanced medical image segmentation (Swin-Unetr) by~\cite{hatamizadeh2021swin}, and (5) an approach to semantic segmentation based on the Vision Transformer (Segmenter) by~\cite{strudel2021segmenter}. All of the parameter settings are followed by original paper.

\begin{figure*}[t!]
\begin{center}
 \includegraphics[width=0.8\linewidth]{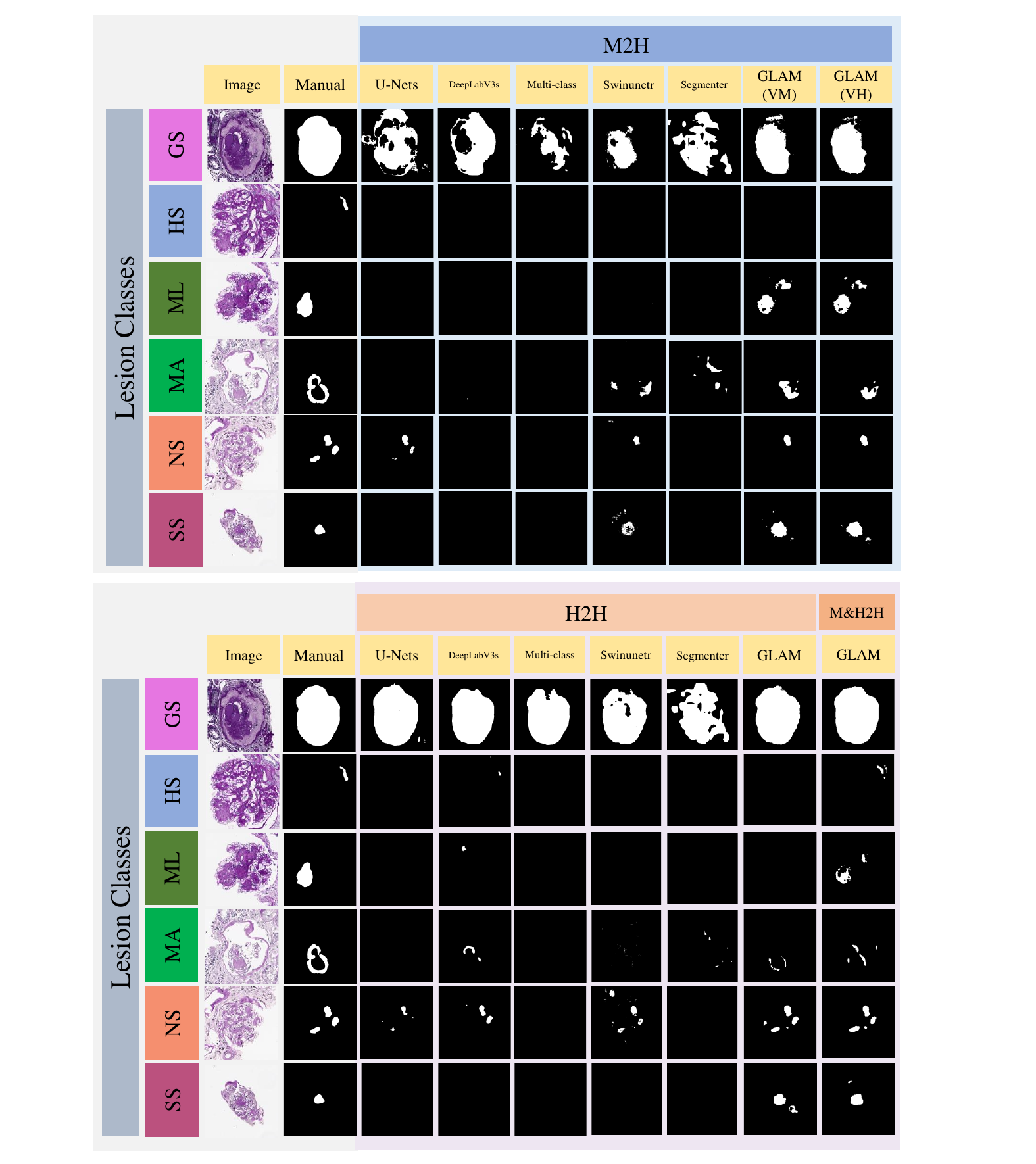}
\end{center}
   \caption{\textbf{Qualitative Results of different methods on human glomerular lesions.} This figure displays the qualitative outcomes of various segmentation methods for six classes of glomerular lesions. The first column features the original, unannotated images, while the second column shows the manual segmentation results. Subsequent columns belong to three section: "Mouse-to-human","Human-to-human" and "Mouse\&Human-to-human", where VM refers to model selection utilizing validation results from mouse data, whereas VH indicates model selection based on validation results from human data. The bold mark indicates the best performance.}
  \label{fig:Quali_hum}
\end{figure*} 
 
\begin{table*}[h!]
% \begin{tabular}{llllllll}
%\centering
\caption{Performance of different hman-to-human (H2H) models trained with human samples on human glomerular lesion segmentation and mouse\&human-to-human (M\&H2H) model trained with hybrid samples . Dice similarity coefficient (\%, the higher, the better), Hausdorff Distance (Micron unit, the lower, the better), and Mean Surface Distance (micron unit, the lower, the better) are used for evaluation. The bold mark indicates the best performance.}

\begin{center}
\begin{adjustbox}{width=1.0\textwidth}
\begin{tabular}{l|cccccccccccc}
\toprule
\multirow{2}{0.8in}{Method} & \multicolumn{3}{c}{Global sclerosis} & \multicolumn{3}{c}{Hyalinosis} & \multicolumn{3}{c}{Mesangial lysis} & \multicolumn{3}{c}{Microaneurysm}\\
\cmidrule(lr){2-4}
\cmidrule(lr){5-7}
\cmidrule(lr){8-10}
\cmidrule(lr){11-13}
& Dice & HD & MSD & Dice & HD & MSD	 & Dice& HD & MSD
 & Dice & HD & MSD\\
\midrule
U-Nets H2H & 91.3 & 120.7 & 23.3 & 49.9 & 391.0 & 355.7 & 49.2 & 455.2 & 375.4 & 49.4 & 422.9 & 347.8\\
DeepLabV3s H2H & \textbf{95.0} & \textbf{69.6} & 15.6 & 54.1 & 313.5 & 241.7 & 54.1 & 336.7 & \textbf{200.6} & \textbf{55.7} & \textbf{235.0} & 137.3\\
Multi-class H2H & 92.6 & 103.8 & 17.7 & 49.9  & 391.0 & 355.7 & 49.2 & 521.6 & 441.8 & 49.4 & 422.9 & 347.8\\
Swinunetr H2H & 90.4 & 131.1 & 26.1 & 49.9 & 314.8 & 285.3 & 49.2 & 382.1 & 273.6 & 49.5 & 336.7 & 252.5\\
Segmenter H2H & 83.3 & 151.0 & 36.6 & 50.2 & 287.1 & 233.6 & 49.2 & 483.8 & 406.6 & 50.5 & 262.6 & \textbf{136.1}\\
\bottomrule
GLAM H2H & 94.3 & 79.8 & \textbf{13.9} & 49.9  & 391.0 & 355.7 & 49.2 & 521.6 & 441.8 & 50.5 & 368.6 & 294.7\\
\bottomrule
GLAM M\&H2H & 94.1 & 76.3 & 14.3 & \textbf{57.8} & \textbf{181.5} & \textbf{85.5} & \textbf{54.6} & \textbf{284.7} & 213.5 & 51.0 & 241.7 & 162.4 \\
\bottomrule

\end{tabular}
\end{adjustbox}
%\begin{adjustbox}{width=1.0\textwidth}
\begin{tabular}{l|cccccccccccc}
\toprule
\multirow{2}{0.8in}{Method} & \multicolumn{3}{c}{Nodular sclerosis
} & \multicolumn{3}{c}{Segmental sclerosis} & \multicolumn{3}{c}{Average}\\
\cmidrule(lr){2-4}
\cmidrule(lr){5-7}
\cmidrule(lr){8-10}
\cmidrule(lr){11-13}
& Dice & HD & MSD & Dice & HD & MSD	 & Dice& HD & MSD\\
\midrule
U-Nets H2H & 67.0 & 158.6 & 55.8 & 47.8 & 492.8 & 399.9  & 59.1 & 340.2 & 259.7\\
DeepLabV3s H2H & 70.9 & 151.4 & 58.9 & 52.1 & 331.0 & 220.6 & 63.7 & 239.5 & 145.8 \\
Multi-class H2H & 48.4  & 490.3 & 371.2 & 47.8 & 492.8 & 399.9 & 56.2 & 403.7 & 322.3 \\
Swinunetr H2H & 65.0  & 183.5 & 49.2 & 47.8 & 492.8 & 399.9 & 58.6 & 306.8 & 214.4 \\
Segmenter H2H & 49.2 & 310.3 & 202.0 & 45.0 & 283.0 & 177.5 & 54.6 & 296.3 & 198.7 \\
\bottomrule
GLAM H2H & 76.5  & 152.2 & 38.8 & 68.5 & \textbf{181.2} & \textbf{57.0} & 64.8 & 282.4 & 200.3\\
\bottomrule
GLAM M\&H2H & \textbf{76.8} & \textbf{145.9} & \textbf{37.0} & \textbf{70.1} & 200.9 & 63.5 &\textbf{ 67.3} & \textbf{188.5} &\textbf{ 96.0} \\
\bottomrule

\end{tabular}
% \end{adjustbox}
\end{center}
\label{table:H2H_result}
\end{table*}

\section{Results}

\subsection{Mouse-to-Human Zero-shot Transfer Learning}
To evaluate the model's performance in adapting from mouse data to human data, we conducted a zero-shot transfer learning experiment. As shown in Table \ref{table:M2H_result}, we assessed the predictive accuracy of models trained exclusively on mouse samples across four lesion categories in human subjects. Figure \ref{fig:Quali_hum} provides a qualitative comparison of the performance of different methods in adapting mouse models for prediction tasks on the human dataset.

\subsection{Mouse \& Human-to-Human Transfer Learning}
We then moved to a more conventional supervised learning scenario, where the model was trained using human samples only. The results, presented in Table \ref{table:H2H_result}, include performance metrics for six lesion classes across the entire human dataset.

Finally, we explored the model's ability to generalize across species by training it on a combined dataset of both mouse and human samples. This hybrid training approach aims to leverage the strengths of both datasets to improve prediction accuracy on human samples. The performance of this model, detailed in Table \ref{table:H2H_result}, shows metrics for six lesion classes across the entire human dataset. 

% According to our experimental results, the GLAM MH2H model not only successfully transfers learning across species but also outperforms most baseline human-to-human models in predicting different lesion classes. This highlights the potential benefits of integrating cross-species data to enhance model performance in complex prediction tasks. 

\section{Conclusion}
In this study, we introduced a deep learning approach for fine-grained segmentation of human kidney lesions by adapting a mouse model, focusing on mouse-to-human transfer learning for glomerular lesions. For zero-shot transfer learning, we developed GLAM zero-shot models, designed to generate precise masks for lesion segmentation. Experimental results indicated that our zero-shot model surpassed baseline methods in segmenting glomerular lesions. For transfer learning with a hybrid dataset, we trained models using both mouse and human samples to predict on a human dataset in order to leverages mouse samples as supplementary support to enhance performance in predicting human lesions. The results demonstrated that the hybrid model generally outperformed baseline methods and achieved satisfactory performance in predicting human datasets.

%\newpage
\appendix
\section{Mouse-to-Mouse Supervised Learning}
%\subsection{Mouse-to-Mouse Supervised Learning}
For evaluating the performance of prediction on mouse, under mouse-to-mouse (M2M) scenario, the model was trained using the mouse training dataset over 200 epochs and its performance was assessed on the mouse testing dataset. As shown in Table \ref{table:M2M_result},the performance metrics for the segmentation of each glomerular lesion class over the entire mouse dataset. 

\begin{table*}[h!]
% \begin{tabular}{llllllll}
%\centering
\caption{Performance of different mouse-to-mouse (M2M) models trained with mouse samples on mouse glomerular lesion segmentation. Dice similarity coefficient (\%, the higher, the better), Hausdorff Distance (Micron unit, the lower, the better), and Mean Surface Distance (micron unit, the lower, the better) are used for evaluation. The bold mark indicates the best performance.}

\begin{center}
\begin{adjustbox}{width=1.0\textwidth}
\begin{tabular}{l|cccccccccccc}
\toprule
\multirow{2}{0.8in}{Method} & \multicolumn{3}{c}{Global sclerosis} & \multicolumn{3}{c}{Hyalinosis} & \multicolumn{3}{c}{Mesangial lysis} & \multicolumn{3}{c}{Microaneurysm}\\
\cmidrule(lr){2-4}
\cmidrule(lr){5-7}
\cmidrule(lr){8-10}
\cmidrule(lr){11-13}
& Dice & HD & MSD & Dice & HD & MSD	 & Dice& HD & MSD
 & Dice & HD & MSD\\
\midrule
U-Nets M2M & 81.5 & 169.9 & 43.6 & 70.5 & 156.0 & 68.6 & 49.4 & 432.0 & 394.5 & 71.6 & 145.8 & 59.4 \\
DeepLabV3s M2M & 87.9 &\textbf{87.4} & \textbf{21.9} & 68.5 & 128.1 & 91.5 & 52.0 & 377.2 & 343.3 & 69.5 & \textbf{110.0} & 46.3 \\
Multi-class M2M & 87.6 & 123.0 & 30.5 & 49.9 & 391.9 & 368.5 & 49.4 & 432.0 & 394.5 & 49.0 & 443.9 & 371.6\\
Swinunetr M2M & 84.3 & 128.1 & 33.5 & 73.3 & 90.6 & 45.1 & 50.8 & 164.4 & 88.8 & 70.5 & 136.9 & \textbf{36.1} \\
Segmenter M2M & 79.8 & 141.0 & 34.3 & 50.8 & 341.3 & 314.2 & 50.2 & 236.0 & 183.2 & 66.8 & 218.4 & 64.3 \\
\bottomrule
GLAM M2M & \textbf{90.0} & 104.8 & 25.3 & \textbf{76.6} & \textbf{81.2} & \textbf{29.5} & \textbf{62.0} & \textbf{148.3} & \textbf{79.3} & \textbf{76.3} & 132.2 & 37.1 \\
\bottomrule

\end{tabular}
\end{adjustbox}
%\begin{adjustbox}{width=1.0\textwidth}
\begin{tabular}{l|cccccccccccc}
\toprule
\multirow{2}{0.8in}{Method} & \multicolumn{3}{c}{Nodular sclerosis
} & \multicolumn{3}{c}{Segmental sclerosis} & \multicolumn{3}{c}{Average}\\
\cmidrule(lr){2-4}
\cmidrule(lr){5-7}
\cmidrule(lr){8-10}
\cmidrule(lr){11-13}
& Dice & HD & MSD & Dice & HD & MSD	 & Dice& HD & MSD\\
\midrule
U-Nets M2M & 62.3 & 205.4 & 135.0 & 47.4 & 475.0 & 372.0 & 63.8 & 264.0 & 178.9\\
DeepLabV3s M2M & 70.1 & 126.7 & 62.9 & 66.0 & 143.9 & 54.8 & 69.0 & 162.2 & 103.5 \\
Multi-class M2M  & 49.2 & 438.6 & 369.2 & 47.4 & 475.0 & 371.9 & 55.4 & 384.1 & 317.7 \\
Swinunetr M2M & 66.8 & \textbf{108.2} & \textbf{32.4} & 49.3 & \textbf{128.7} & 47.9 & 65.8 & 126.2 & 47.3 \\
Segmenter M2M & 56.5 & 166.0 & 59.0 & 51.5 & 202.2 & 80.1  & 59.3 & 217.5 & 122.5 \\
\bottomrule
GLAM M2M &  \textbf{77.7} & 108.3 & 37.2 & \textbf{78.2} & 145.9 & \textbf{39.6} & \textbf{76.8} & \textbf{120.1} & \textbf{41.3}\\
\bottomrule

\end{tabular}
%\end{adjustbox}
\end{center}
\label{table:M2M_result}
\end{table*}

\begin{figure*}[bth]
\begin{center}
 \includegraphics[width=0.8\linewidth]{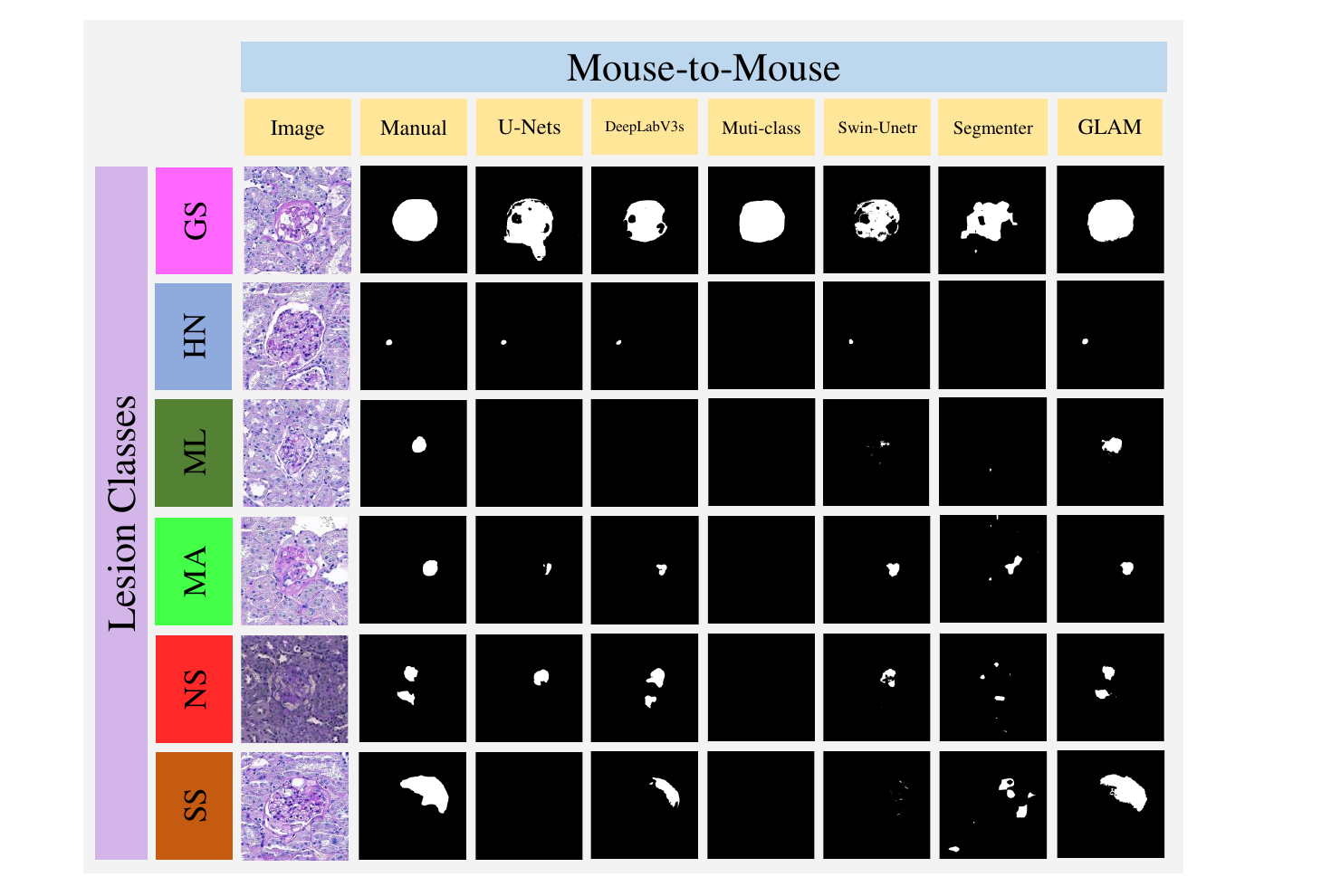}
\end{center}
   \caption{\textbf{Qualitative Results of different methods on six mice lesions for diabetic nephropathy.} This figure displays the qualitative outcomes of various segmentation methods for six classes of glomerular lesions. The first column features the original, unannotated images, while the second column shows the manual segmentation results. Subsequent columns are dedicated to the results from baseline models and the GLAM Mouse-to-Mouse approach, respectively.}
  \label{fig:Quali_mice}
\end{figure*} 

Fig.\ref{fig:Quali_mice} presents the qualitative result about the performance of different methods on the mouse lesion dataset. The experimental results show that GLAM M2M as a single multi-label model can achieve better performance on prediction classes of glomerular lesions than baseline methods. such as CNNs-based.DeepLabV3 and Transformers-based (i.e.Swin-unetr).

\acknowledgments
This research was supported by NIH R01DK135597(Huo), DoD HT9425-23-1-0003(HCY), NIH NIDDK DK56942 \\(ABF). This work was also supported by Vanderbilt Seed Success Grant, Vanderbilt Discovery Grant, and VISE Seed Grant. This project was supported by The Leona M. and Harry B. Helmsley Charitable Trust grant G-1903-03793 and G-2103-05128. This research was also supported by NIH grants R01EB033385, R01DK132338, REB017230, R01MH125931, and NSF 2040462. We extend gratitude to NVIDIA for their support by means of the NVIDIA hardware grant. This works was also supported by NSF NAIRR Pilot Award NAIRR240055.

% References
\bibliography{report} % bibliography data in report.bib

\begin{thebibliography}{10}

\bibitem{isensee2021nnu}
Isensee, F., Jaeger, P.~F., Kohl, S.~A., Petersen, J., and Maier-Hein, K.~H., ``nnu-net: a self-configuring method for deep learning-based biomedical image segmentation,'' {\em Nature methods}~{\bf 18}(2),  203--211 (2021).

\bibitem{huang2017densely}
Huang, G., Liu, Z., Van Der~Maaten, L., and Weinberger, K.~Q., ``Densely connected convolutional networks,'' in [{\em Proceedings of the IEEE conference on computer vision and pattern recognition}{\nolinebreak\hspace{0.1em}]},   4700--4708 (2017).

\bibitem{dolz2018hyperdense}
Dolz, J., Gopinath, K., Yuan, J., Lombaert, H., Desrosiers, C., and Ayed, I.~B., ``Hyperdense-net: a hyper-densely connected cnn for multi-modal image segmentation,'' {\em IEEE transactions on medical imaging}~{\bf 38}(5),  1116--1126 (2018).

\bibitem{jiang2021deep}
Jiang, L., Chen, W., Dong, B., Mei, K., Zhu, C., Liu, J., Cai, M., Yan, Y., Wang, G., Zuo, L., et~al., ``A deep learning-based approach for glomeruli instance segmentation from multistained renal biopsy pathologic images,'' {\em The American Journal of Pathology}~{\bf 191}(8),  1431--1441 (2021).

\bibitem{han2023fastcellpose}
Han, Y., Zhang, Z., Li, Y., Fan, G., Liang, M., Liu, Z., Nie, S., Ning, K., Luo, Q., and Yuan, J., ``Fastcellpose: A fast and accurate deep-learning framework for segmentation of all glomeruli in mouse whole-kidney microscopic optical images,'' {\em Cells}~{\bf 12}(23),  2753 (2023).

\bibitem{ostergaard2020automated}
{\O}stergaard, M.~V., Sembach, F.~E., Skytte, J.~L., Roostalu, U., Secher, T., Overgaard, A., Fink, L.~N., Vrang, N., Jelsing, J., and Hecksher-S{\o}rensen, J., ``Automated image analyses of glomerular hypertrophy in a mouse model of diabetic nephropathy,'' {\em Kidney360}~{\bf 1}(6),  469 (2020).

\bibitem{janowczyk2016deep}
Janowczyk, A. and Madabhushi, A., ``Deep learning for digital pathology image analysis: A comprehensive tutorial with selected use cases,'' {\em Journal of pathology informatics}~{\bf 7}(1),  29 (2016).

\bibitem{komura2019machine}
Komura, D. and Ishikawa, S., ``Machine learning approaches for pathologic diagnosis,'' {\em Virchows Archiv}~{\bf 475},  131--138 (2019).

\bibitem{gadermayr2019cnn}
Gadermayr, M., Dombrowski, A.-K., Klinkhammer, B.~M., Boor, P., and Merhof, D., ``Cnn cascades for segmenting sparse objects in gigapixel whole slide images,'' {\em Computerized Medical Imaging and Graphics}~{\bf 71},  40--48 (2019).

\bibitem{esteva2019guide}
Esteva, A., Robicquet, A., Ramsundar, B., Kuleshov, V., DePristo, M., Chou, K., Cui, C., Corrado, G., Thrun, S., and Dean, J., ``A guide to deep learning in healthcare,'' {\em Nature medicine}~{\bf 25}(1),  24--29 (2019).

\bibitem{wang2019pathology}
Wang, S., Yang, D.~M., Rong, R., Zhan, X., and Xiao, G., ``Pathology image analysis using segmentation deep learning algorithms,'' {\em The American journal of pathology}~{\bf 189}(9),  1686--1698 (2019).

\bibitem{kamnitsas2017efficient}
Kamnitsas, K., Ledig, C., Newcombe, V.~F., Simpson, J.~P., Kane, A.~D., Menon, D.~K., Rueckert, D., and Glocker, B., ``Efficient multi-scale 3d cnn with fully connected crf for accurate brain lesion segmentation,'' {\em Medical image analysis}~{\bf 36},  61--78 (2017).

\bibitem{souza2023mouse}
Souza, L., Silva, J., Chagas, P., Duarte, A., Santos, W. L.-d., and Oliveira, L., ``Mouse-to-human transfer learning for glomerulus segmentation,'' {\em Computer Methods in Biomechanics and Biomedical Engineering: Imaging \& Visualization} ,  1--10 (2023).

\bibitem{yang2022glomerular}
Yang, C.-K., Lee, C.-Y., Wang, H.-S., Huang, S.-C., Liang, P.-I., Chen, J.-S., Kuo, C.-F., Tu, K.-H., Yeh, C.-Y., and Chen, T.-D., ``Glomerular disease classification and lesion identification by machine learning,'' {\em biomedical journal}~{\bf 45}(4),  675--685 (2022).

\bibitem{saikia2023mlp}
Saikia, F.~N., Iwahori, Y., Suzuki, T., Bhuyan, M., Wang, A., and Kijsirikul, B., ``Mlp-unet: Glomerulus segmentation,'' {\em IEEE Access}  (2023).

\bibitem{deng2022single}
Deng, R., Liu, Q., Cui, C., Asad, Z., Huo, Y., et~al., ``Single dynamic network for multi-label renal pathology image segmentation,'' in [{\em International Conference on Medical Imaging with Deep Learning}{\nolinebreak\hspace{0.1em}]},   304--314, PMLR (2022).

\bibitem{chen2017fast}
Chen, Q., Xu, J., and Koltun, V., ``Fast image processing with fully-convolutional networks,'' in [{\em Proceedings of the IEEE International Conference on Computer Vision}{\nolinebreak\hspace{0.1em}]},   2497--2506 (2017).

\bibitem{zhang2021dodnet}
Zhang, J., Xie, Y., Xia, Y., and Shen, C., ``Dodnet: Learning to segment multi-organ and tumors from multiple partially labeled datasets,'' in [{\em Proceedings of the IEEE/CVF conference on computer vision and pattern recognition}{\nolinebreak\hspace{0.1em}]},   1195--1204 (2021).

\bibitem{zhu2017unpaired}
Zhu, J.-Y., Park, T., Isola, P., and Efros, A.~A., ``Unpaired image-to-image translation using cycle-consistent adversarial networks,'' in [{\em Proceedings of the IEEE international conference on computer vision}{\nolinebreak\hspace{0.1em}]},   2223--2232 (2017).

\bibitem{ronneberger2015u}
Ronneberger, O., Fischer, P., and Brox, T., ``U-net: Convolutional networks for biomedical image segmentation,'' in [{\em Medical Image Computing and Computer-Assisted Intervention--MICCAI 2015: 18th International Conference, Munich, Germany, October 5-9, 2015, Proceedings, Part III 18}{\nolinebreak\hspace{0.1em}]},   234--241, Springer (2015).

\bibitem{lutnick2019integrated}
Lutnick, B., Ginley, B., Govind, D., McGarry, S.~D., LaViolette, P.~S., Yacoub, R., Jain, S., Tomaszewski, J.~E., Jen, K.-Y., and Sarder, P., ``An integrated iterative annotation technique for easing neural network training in medical image analysis,'' {\em Nature machine intelligence}~{\bf 1}(2),  112--119 (2019).

\bibitem{gonzalez2018multi}
Gonz{\'a}lez, G., Washko, G.~R., and San Jos{\'e}~Est{\'e}par, R., ``Multi-structure segmentation from partially labeled datasets. application to body composition measurements on ct scans,'' in [{\em International Workshop on Reconstruction and Analysis of Moving Body Organs}{\nolinebreak\hspace{0.1em}]},   215--224, Springer (2018).

\bibitem{hatamizadeh2021swin}
Hatamizadeh, A., Nath, V., Tang, Y., Yang, D., Roth, H.~R., and Xu, D., ``Swin unetr: Swin transformers for semantic segmentation of brain tumors in mri images,'' in [{\em International MICCAI Brainlesion Workshop}{\nolinebreak\hspace{0.1em}]},   272--284, Springer (2021).

\bibitem{strudel2021segmenter}
Strudel, R., Garcia, R., Laptev, I., and Schmid, C., ``Segmenter: Transformer for semantic segmentation,'' in [{\em Proceedings of the IEEE/CVF international conference on computer vision}{\nolinebreak\hspace{0.1em}]},   7262--7272 (2021).

\end{thebibliography}
\bibliographystyle{spiebib} % makes bibtex use spiebib.bst

\end{document}